# A Backtracking-Based Algorithm for Computing Hypertree-Decompositions[*]

Georg Gottlob[1] and Marko Samer[2]

[1] Computing Laboratory
Oxford University, UK
`georg.gottlob@comlab.ox.ac.uk`
[2] Department of Computer Science
Durham University, UK
`marko.samer@durham.ac.uk`

**Abstract.** Hypertree decompositions of hypergraphs are a generalization of tree decompositions of graphs. The corresponding hypertree-width is a measure for the cyclicity and therefore tractability of the encoded computation problem. Many NP-hard decision and computation problems are known to be tractable on instances whose structure corresponds to hypergraphs of bounded hypertree-width. Intuitively, the smaller the hypertree-width, the faster the computation problem can be solved. In this paper, we present the new backtracking-based algorithm `det-`$k$`-decomp` for computing hypertree decompositions of small width. Our benchmark evaluations have shown that `det-`$k$`-decomp` significantly outperforms `opt-`$k$`-decomp`, the only exact hypertree decomposition algorithm so far. Even compared to the best heuristic algorithm, we obtained competitive results as long as the hypergraphs are not too large.

## 1 Introduction

Since many important problems in computer science are intractable in general, it is a competitive task to identify tractable subclasses of such problems which can be solved efficiently. One approach to do this is to restrict the structure of a problem represented as graph or hypergraph. For example, the structure of instances of the constraint satisfaction problem CSP and the equivalent Boolean conjunctive query problem BCQ can be naturally encoded by hypergraphs. A hypergraph is a generalization of a graph where each edge connects a set of vertices. Gottlob et al. [3] have shown that in analogy to tree-width of graphs, the hypertree-width of hypergraphs is an appropriate measure for the cyclicity and therefore the tractability of the corresponding computation problems. The hypertree-width of a hypergraph is defined as the minimum width over all decompositions of the hypergraph in a way called *hypertree decomposition*. Roughly speaking, the smaller the width of a hypertree decomposition, the faster the corresponding problem can be solved. However, deciding whether there exists a hypertree decomposition of width at most $k$ is NP-complete in general.

[*] Research supported by the Austrian Science Fund (FWF), P17222-N04.

Gottlob et al. [3] have shown that for fixed $k$, the problem of deciding whether there exists a hypertree decomposition of width at most $k$ is in LOGCFL, i.e., it can be solved in polynomial time and is highly parallelizable. Moreover, they presented the alternating algorithm `k-decomp`, which constructs a hypertree decomposition of minimal width less than or equal to $k$ (if such a decomposition exists). Another algorithm for computing hypertree decompositions of minimal width less than or equal to $k$ is `opt-k-decomp` [2, 7, 9]. Up to now, `opt-k-decomp` has been the only implementable exact polynomial-time algorithm for constructing $k$-bounded hypertree decompositions. However, `opt-k-decomp` has an important disadvantage: although polynomial, it needs a huge amount of memory and time even for small hypergraphs. This basic problem remains even after improvements like redundancy elimination as investigated by Harvey and Ghose [5]. Thus, recent research focuses on heuristic approaches for constructing hypertree decompositions of small but not necessarily minimal width. One of the practically most successful algorithms of this kind is due to McMahan [8], who combined well-known tree decomposition and set cover heuristics in order to construct hypertree decompositions. This approach, however, has the disadvantage that the computation cannot be focused on hypertrees of width smaller than some upper bound $k$, i.e., the algorithm returns a hypertree decomposition whose width may be much larger than the minimal one although there is often time left to improve this result.

In this paper, we consider a combination of exact and heuristic approaches in the sense that we restrict the search space by a fixed upper bound $k$ and apply heuristics to accelerate the search for a hypertree decomposition of width at most $k$ (but not necessarily the minimal one). Such an approach has the following advantages: the search space and the width of the resulting hypertree decomposition can be restricted by $k$. In particular, this means that the algorithm is able to find a hypertree decomposition of minimal width by setting $k$ small enough. On the other hand, if $k$ is not minimal, the algorithm may construct a hypertree decomposition of non-minimal width but within a reasonable amount of time since the solution space becomes larger. In principle, such an approach allows us to control the tradeoff between the resulting hypertree width and the required computation time. Thus, it combines the advantages of exact and heuristic algorithms while it minimizes their disadvantages.

We implemented our algorithm `det-k-decomp` based on these insights. Our experimental evaluations have shown that it performs much better than `opt-k-decomp`. In particular, it needs only a small amount of memory and is much faster than `opt-k-decomp`. Moreover, it is often the case that `det-k-decomp` finds hypertree decompositions of width smaller than those obtained by McMahan's heuristic algorithm. Finally, there is a further advantage of our algorithm: due to its top-down nature of decomposing a hypergraph into sub-hypergraphs and decomposing the sub-hypergraphs into sub-sub-hypergraphs etc. it can in principle be implemented for parallel execution in order to increase its performance even further.

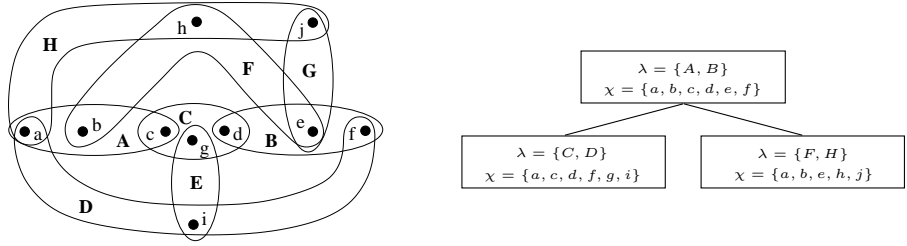

**Fig. 1.** Example of a hypergraph and its hypertree decomposition of width 2

This paper is organized as follows: In Section 2, we give the basic definitions used in this paper. Afterwards, in Section 3, we describe the alternating algorithm `k-decomp`. Then, in Section 4, we introduce our new algorithm `det-k-decomp` and prove its correctness and polynomial runtime. In Section 5, we present our evaluation results. Finally, we conclude in Section 6.

## 2 Preliminaries

A *hypergraph* $H$ is a tuple $(V, E)$, where $V$ is a set of vertices (variables) and $E \subseteq 2^V \setminus \{\emptyset\}$ is a set of hyperedges (atoms); w.l.o.g., we assume that $V \setminus \bigcup E = \emptyset$. We define $vertices(H) = V$ and $edges(H) = E$. A *hypertree for a hypergraph $H$* is a triple $(T, \chi, \lambda)$, where $T = (V, E)$ is a tree and $\chi : V \to 2^{vertices(H)}$ and $\lambda : V \to 2^{edges(H)}$ are labeling functions. We define $vertices(T) = V$ and refer to the vertices of $T$ as "nodes" to avoid confusion with the vertices of $H$. If $T' = (V', E')$ is a subtree of $T$, we define $\chi(T') = \bigcup_{p \in V'} \chi(p)$. We denote the root of $T$ by $root(T)$, and for every $p \in vertices(T)$ we denote the subtree of $T$ rooted at $p$ by $T_p$.

A *hypertree decomposition* of a hypergraph $H$ is a hypertree $(T, \chi, \lambda)$ for $H$ satisfying the following four conditions [3]:

1. $\forall e \in edges(H) \ \exists p \in vertices(T) : e \subseteq \chi(p)$,
2. $\forall v \in vertices(H)$: the set $\{p \in vertices(T) \,|\, v \in \chi(p)\}$ induces a (connected) subtree of $T$,
3. $\forall p \in vertices(T) : \chi(p) \subseteq \bigcup \lambda(p)$, and
4. $\forall p \in vertices(T) : \bigcup \lambda(p) \cap \chi(T_p) \subseteq \chi(p)$.

The *width of a hypertree decomposition* $(T, \chi, \lambda)$ is given by $\max_{p \in vertices(T)} |\lambda(p)|$, and the *hypertree-width of a hypergraph* is the minimum width over all its hypertree decompositions. Fig. 1 shows an example of a hypergraph (on the left) and its hypertree decomposition of width 2 (on the right).

Let $H = (V, E)$ be a hypergraph. A path between two vertices $x, y \in V$ is a sequence of vertices $x = v_1, v_2, \ldots, v_{k-1}, v_k = y$ such that for all pairs $v_i, v_{i+1}$ there exists $e \in E$ with $v_i, v_{i+1} \in e$. Now, let $W \subseteq V$. A set $V' \subseteq V$ is $[W]$-connected if for all $x, y \in V'$ there exists a path between $x$ and $y$ not going through vertices in $W$. A $[W]$-*component* is a maximal $[W]$-connected non-empty set of vertices $V' \subseteq V \setminus W$. We call the set $W$ in this context a *separator*. Note that both, components and separators, can also be represented by the hyperedges covering the corresponding vertices, which we will do in this paper.

## 3  The Algorithm $k$-decomp

In this section, we describe the alternating algorithm $k$-decomp introduced by Gottlob et al. [3]. This algorithm checks non-deterministically whether there exists a hypertree decomposition of a hypergraph of width at most $k$. By using this algorithm, Gottlob et al. proved that the problem of deciding whether a hypergraph has $k$-bounded hypertree-width is in LOGCFL, a very low polynomial-time complexity class which is contained in $NC_2$ and which thus consists of highly parallelizable problems. The algorithm works in a top-down manner by guessing in each step the $\lambda$-labels and checking two conditions. The $\chi$-labels and sub-hypergraphs can then be computed deterministically. A game-theoretic intuition for such an approach can also be found in [4].

The algorithm $k$-decomp shown in Algorithms 1 and 2 is a notational variant of the algorithm presented in [3]. A proof of equivalence of the two variants is given in the appendix. The main procedure $k$-decomp in Algorithm 1 expects a hypergraph `HGraph` consisting of a set of vertices `vertices(HGraph)` and a set of hyperedges `edges(HGraph)` as parameter. It calls the recursive procedure $k$-decomposable in Algorithm 2, which returns a hypertree decomposition of width at most $k$ if it exists and `NULL` otherwise. The parameters of $k$-decomposable are a set `OldSep` of hyperedges which were chosen as separator in the previous run and a set `Edges` of hyperedges which represents a component w.r.t. `OldSep`, i.e., which represents the current sub-hypergraph that has to be decomposed. In line 1, a set `Separator` of hyperedges of size at most $k$ is guessed. In lines 2 to 4 it is then checked whether `Separator` is indeed a separator, i.e., whether it decomposes the current sub-hypergraph into several (at least one) smaller components and whether the conditions of a hypertree decomposition are satisfied. In line 6, the procedure `separate` computes these components and stores them in `Components`. Afterwards, in lines 7 to 15, the hypertree decompositions of the components are recursively computed and stored in `Subtrees`. Line 16 computes the $\chi$-labels of the current hypertree node in such a way that the conditions of a hypertree decomposition are satisfied. Finally, the procedure `getHTNode` in line 17 constructs the current hypertree node consisting of the $\lambda$-labels represented by `Separator`, the $\chi$-labels represented by `Chi`, and the subtrees stored in `Subtrees`.

For example, when applying $k$-decomp to the hypergraph in Fig. 1, $k$-decomposable is called with parameters `Edges` $= \{A, B, C, D, E, F, G, H\}$ and `OldSep` $= \emptyset$. In the first run, we guess `Separator` $= \{A, B\}$ which results in `Components` $= \{\{C, D, E\}, \{F, G, H\}\}$. Consequently, there are two recursive calls: the first one uses parameters `Edges` $= \{C, D, E\}$ and `OldSep` $= \{A, B\}$, and the second one uses `Edges` $= \{F, G, H\}$ and `OldSep` $= \{A, B\}$. In the fol-lowing recursion level, we guess `Separator` $= \{C, D\}$ and `Separator` $= \{F, H\}$ respectively. It is easy to see that no further recursive calls are necessary since both sub-hypergraphs are then completely decomposed. The resulting hypertree decomposition is shown in Fig. 1.

---

**Algorithm 1** *k-decomp(HGraph)*

---

1  *HTree* := *k-decomposable(edges(HGraph), ∅)*;
2  **return** *HTree*;

---

**Algorithm 2** *k-decomposable(Edges, OldSep)*

---

1  **guess** *Separator* ⊆ *edges(HGraph)* **such that** |*Separator*| ≤ *k*;
2  **check** that the following two conditions hold:
3      $\bigcup Edges \cap \bigcup OldSep \subseteq \bigcup Separator$;
4      *Separator* ∩ *Edges* ≠ ∅;
5  **if** one of these checks fails **then return** *NULL*;
6  *Components* := *separate(Edges, Separator)*;
7  *Subtrees* := ∅;
8  **for each** *Comp* ∈ *Components* **do**
9      *HTree* := *k-decomposable(Comp, Separator)*;
10     **if** *HTree* = *NULL* **then**
11         **return** *NULL*;
12     **else**
13         *Subtrees* := *Subtrees* ∪ {*HTree*};
14     **endif**
15 **endfor**
16 *Chi* := ($\bigcup Edges \cap \bigcup OldSep$) ∪ $\bigcup(Separator \cap Edges)$;
17 *HTree* := *getHTNode(Separator, Chi, Subtrees)*;
18 **return** *HTree*;

---

As it was shown in [3], `k-decomp` can be implemented on a logspace ATM having polynomially bounded tree-size. Hence, deciding whether a hypergraph has $k$-bounded hypertree-width is in LogCFL and thus in P. However, since `k-decomp` is a non-deterministic algorithm, this is only of theoretical interest and cannot be implemented. The only deterministic algorithm with polynomial runtime for computing $k$-bounded hypertree decompositions published so far is `opt-k-decomp` [2, 7, 9]. This algorithm, however, has a basic drawback: due to its bottom-up approach, it needs a huge amount of memory even for small examples. So it is practically only applicable to hypergraphs of very small size and simple structure. Our aim in the following section is therefore to transform `k-decomp` into a deterministic algorithm with polynomial runtime which performs better than `opt-`$k$`-decomp`. We call our resulting algorithm `det-`$k$`-decomp`.

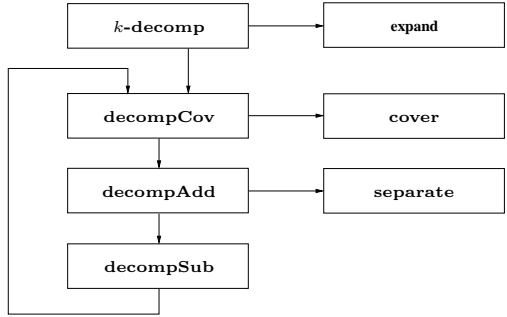

**Fig. 2.** Dependency graph between procedures called by $k$-decomp

## 4 The Algorithm det-$k$-decomp

In this section, we present our new algorithm `det-`$k$`-decomp` which we obtain from $k$`-decomp` by replacing the *guess and check*-part in lines 1 to 4 of Algorithm 2 by a backtracking-based search procedure. In order to preserve the polynomial runtime, however, we have to store already visited separators; otherwise, the same subtree could be constructed multiple times which would lead to an exponential runtime. Algorithm 3 represents the main procedure which calls the recursive procedures `decompCov`, `decompAdd`, and `decompSub` described in Algorithms 4 to 6. We divided the recursive part into these three sub-procedures for better readability. Their dependencies are shown in Fig. 2, where an arrow from $A$ to $B$ means that $A$ calls $B$. Moreover, we use three auxiliary procedures:

- `separate` is the same as for $k$`-decomp` described in Section 3 and computes the components of the current sub-hypergraph w.r.t. the chosen separator.
- `cover` selects hyperedges for the separator which are satisfying the first condition of the *check*-step in Algorithm 2; we will describe it later in more detail.
- `expand` is used to complete a "pruned" hypertree decomposition: if the same hypertree node is constructed the second time during our search procedure, we know already if the corresponding components can be decomposed. So we have to prune the hypertree at this node in order to guarantee the polynomial runtime as mentioned above. The procedure `expand` expands such nodes to subtrees after the search process, which can be done in polynomial time.

We are now going to explain the transformation from $k$`-decomp` to `det-`$k$`-decomp` in more detail. The main procedures in Algorithm 1 and Algorithm 3 are almost the same. The only differences are the instantiation of the sets `FailSeps`

**Algorithm 3** *det-k-decomp*(*HGraph*)

1. *FailSeps* := ∅;
2. *SuccSeps* := ∅;
3. *HTree* := *decompCov*(*edges*(*HGraph*), ∅);
4. **if** *HTree* ≠ *NULL* **then**
5.     *HTree* := *expand*(*HTree*);
6. **endif**
7. **return** *HTree*;

---

**Algorithm 4** *decompCov*(*Edges*, *Conn*)

1. **if** |*Edges*| ≤ $k$ **then**
2.     *HTree* := *getHTNode*(*Edges*, $\bigcup$ *Edges*, ∅);
3.     **return** *HTree*;
4. **endif**
5. *BoundEdges* := {$e \in edges(HGraph) \mid e \cap Conn \neq \emptyset$};
6. **for each** *CovSep* ∈ *cover*(*Conn*, *BoundEdges*) **do**
7.     *HTree* := *decompAdd*(*Edges*, *Conn*, *CovSep*);
8.     **if** *HTree* ≠ *NULL* **then**
9.         **return** *HTree*;
10.     **endif**
11. **endfor**
12. **return** *NULL*;

---

**Algorithm 5** *decompAdd*(*Edges*, *Conn*, *CovSep*)

1. *InCovSep* := *CovSep* ∩ *Edges*;
2. **if** *InCovSep* ≠ ∅ **or** $k - |CovSep| > 0$ **then**
3.     **if** *InCovSep* = ∅ **then** *AddSize* := 1 **else** *AddSize* := 0 **endif**;
4.     **for each** *AddSep* ⊆ *Edges* **s.t.** |*AddSep*| = *AddSize* **do**
5.         *Separator* := *CovSep* ∪ *AddSep*;
6.         *Components* := *separate*(*Edges*, *Separator*);
7.         **if** ∀*Comp* ∈ *Components*. ⟨*Separator*, *Comp*⟩ ∉ *FailSeps* **then**
8.             *Subtrees* := *decompSub*(*Components*, *Separator*);
9.             **if** *Subtrees* ≠ ∅ **then**
10.                 *Chi* := *Conn* ∪ $\bigcup$(*InCovSep* ∪ *AddSep*);
11.                 *HTree* := *getHTNode*(*Separator*, *Chi*, *Subtrees*);
12.                 **return** *HTree*;
13.             **endif**
14.         **endif**
15.     **endfor**
16. **endif**
17. **return** *NULL*;

**Algorithm 6** *decompSub*(*Components*, *Separator*)

1  *Subtrees* := ∅;
2  **for each** *Comp* ∈ *Components* **do**
3      *ChildConn* := ⋃ *Comp* ∩ ⋃ *Separator*;
4      **if** ⟨*Separator*, *Comp*⟩ ∈ *SuccSeps* **then**
5          *HTree* := *getHTNode*(*Comp*, *ChildConn*, ∅);
6      **else**
7          *HTree* := *decompCov*(*Comp*, *ChildConn*);
8          **if** *HTree* = *NULL* **then**
9              *FailSeps* := *FailSeps* ∪ {⟨*Separator*, *Comp*⟩};
10             **return** ∅;
11         **else**
12             *SuccSeps* := *SuccSeps* ∪ {⟨*Separator*, *Comp*⟩};
13         **endif**
14     **endif**
15     *Subtrees* := *Subtrees* ∪ {*HTree*};
16 **endfor**
17 **return** *Subtrees*;

and `SuccSeps`, which are used to store already visited separators, and the call of `expand`. If a component could be successfully decomposed, the corresponding separator together with the component is inserted into `SuccSeps`; otherwise it is inserted into `FailSeps`. If a new separator is constructed during the search, it is checked whether it occurs in one of these two sets before its components are tried to be decomposed. This avoids that the same components are decomposed multiple times. Intuitively, since there are at most $\mathcal{O}(n^k)$ separators each of size at most $k$ and each with at most $n = |edges(HGraph)|$ components, the search procedure runs in polynomial time for fixed $k$.

Now, let us consider the recursive part in Algorithms 4 to 6. First note that the second parameter of `decompCov` is a set of vertices instead of a set of hyperedges in $k$-`decomposable`. This is just a simplification since `OldSep` in Algorithm 2 occurs only in the expression ⋃ `Edges` ∩ ⋃ `OldSep`. So we compute this set of "connecting vertices" `Conn` before the recursive call and replace `OldSep` by `Conn`.

The first few lines in Algorithm 4 are a simple optimization: if the current component contains at most $k$ hyperedges, it can be trivially decomposed into a single hypertree-node. Afterwards, in line 5, our determinization of the *guess and check*-part in Algorithm 2 starts. In particular, we compute the set `BoundEdges` of hyperedges which are necessary to satisfy the condition in line 3 of Algorithm 2, which is `Conn` ⊆ ⋃ `Separator`. This condition can only be satisfied by the separator through hyperedges in `BoundEdges`, i.e., through hyperedges containing some vertices in `Conn`. Thus, we can reduce the search space for an appropriate separator by first selecting a subset `CovSep` of at most $k$ hyperedges in `BoundEdges` such that `Conn` ⊆ ⋃ `CovSep`, where `CovSep` ⊆ `Separator`. This selection is done by the procedure `cover` in line 6. In particular, `cover` successively returns all possible selections and we have to loop through all of them. The further decomposition steps are then performed by `decompAdd`, which is called in line 7. If one such decomposition is successful, the resulting hypertree is returned in line 9; otherwise, `NULL` is returned in line 12 after all selections for `CovSep` have been tried. Note that the order of choosing `CovSep` is crucial for the duration of the search process in the case of success. Thus, we use heuristics in `cover` to obtain an appropriate selection order. We will explain these heuristics later in more detail.

Let us now consider the procedure `decompAdd` defined by Algorithm 5. Its third parameter is the set `CovSep` which contains at most $k$ hyperedges. These hyperedges were selected in such a way that the condition in line 3 of Algorithm 2 is satisfied. Since every separator guessed in `k-decomposable` must satisfy this condition and we loop through all possibilities of `CovSep`, we know that our search space is large enough to find all separators which are accepted by `k-decomposable`. This, however, is not necessary. To improve the runtime of our algorithm, we can restrict the search space based on the following insights.

**Definition 1.** *[3] A hypertree decomposition of a hypergraph is in* normal form *if for each $r \in vertices(T)$ and for each child $s$ of $r$ all the following conditions hold:*

1. *there is (exactly) one $[\chi(r)]$-component $C_r$ such that $\chi(T_s) = C_r \cup (\chi(s) \cap \chi(r))$,*
2. *$\chi(s) \cap C_r \neq \emptyset$, where $C_r$ is the $[\chi(r)]$-component satisfying Condition 1, and*
3. *$\bigcup \lambda(s) \cap \chi(r) \subseteq \chi(s)$.*

**Definition 2.** *Let $(T, \chi, \lambda)$ be a hypertree decomposition of a hypergraph and $s \in vertices(T)$. We say $s$ is* minimally labeled *if $s = root(T)$ and $|\lambda(s)| = 1$ or $r$ is the parent of $s$ (with component $C_r$) and it holds that for each $e \in \lambda(s)$:*

1. $\bigcup edges(C_r) \cap \bigcup \lambda(r) \nsubseteq \bigcup(\lambda(s) \setminus \{e\})$; *or*              *(Connectivity)*
2. $(\lambda(s) \setminus \{e\}) \cap edges(C_r) = \emptyset$.                                 *(Monotonicity)*

Definition 2 means that removing any edge from the $\lambda$-labels violates the connectivity or the monotonicity condition.

**Definition 3.** *A hypertree decomposition of a hypergraph is in* strong normal form *if it is in normal form and each vertex is minimally labeled.*

**Lemma 1.** *For each $k$-width hypertree decomposition of a hypergraph $H$ there exists a $k$-width hypertree decomposition of $H$ in strong normal form.*

*Proof.* First note that we can assume w.l.o.g. that our hypertree decomposition $(T, \chi, \lambda)$ is in normal form [3]. We will show the lemma by an inductive argument on the size of the $\chi$-components. To this aim, let $s \in vertices(T)$ be not minimally labeled and assume that all other vertices on the path from $s$ to $root(T)$ are minimally labeled. Now, we distinguish between two cases as illustrated in Fig. 3: (a) If $s = root(T)$, we create a new vertex $r'$ and add it to $T$ as parent of $s$. We set $\lambda(r') = \{e\}$ for any hyperedge $e \in \lambda(s)$. (b) Otherwise, if $s \neq root(T)$, let $r$ be the parent of $s$. We create a new vertex $r'$ and add it to $T$ between $s$ and $r$, that is, $r'$ is the parent of $s$ and $r$ is the parent of $r'$. We set $\lambda(r') = \lambda(s)$ and remove hyperedges from $\lambda(r')$ until any further removal would violate one of the conditions in Definition 2. Finally, we set $\chi(r') = \chi(s) \cap \bigcup \lambda(r')$ and remove all $v \in \chi(r) \setminus \chi(r')$ from $\chi(s')$ for each $s' \in vertices(T_s)$. It is easy to see that the resulting tree is still a hypertree decomposition, however, it is not necessarily in normal form. Since our modifications affect only the subtree rooted at $r'$, we know that this subtree can be transformed into normal form according to [3] without changing the labels of $r'$. Since $C_{r'} \subset C_r$ by the monotonicity condition, our inductive argument applies. □

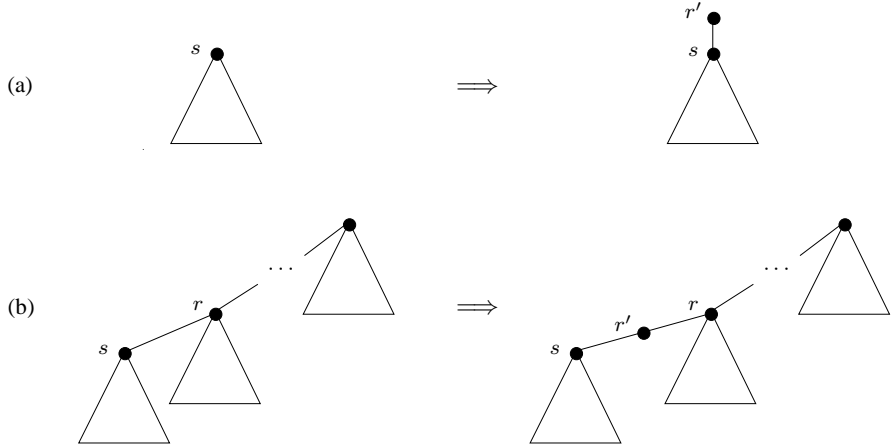

**Fig. 3.** Transformation steps to strong normal form

This means that it suffices to consider the smallest separators satisfying the conditions in Algorithm 2. In particular, for `decompAdd` in Algorithm 5, this implies that it suffices to add at most one hyperedge to `CovSep` such that the condition in line 4 of Algorithm 2 is satisfied. To this aim, `decompAdd` computes in line 1 the intersection between `CovSep` and `Edges`. In line 2 it is then checked if the condition in line 4 of Algorithm 2 can be satisfied at all. In particular, if `InCovSep` is empty then we must add an edge to satisfy the condition. However, if `CovSep` contains already $k$ hyperedges, this is not possible and we have to reject `CovSep`. Otherwise, in line 3 we set `AddSize` to 1 if we must add an edge and to 0 if `CovSep` already satisfies the condition. Then in line 4 we loop through all hyperedges in `Edges` if `AddSize` = 1, that is, if we must add an edge; otherwise, `AddSep` is the empty set and the loop body is executed only once. Finally, our separator is computed in line 5 as the union of `CovSep` and `AddSep`, where `AddSep` contains either a single hyperedge or is empty. In the latter case, `CovSep` satisfies both conditions in Algorithm 2 (recall that it satisfies the first condition by construction). Based on Lemma 1, we know that our computation of `Separator` is a correct determinization of the *guess and check*-part in Algorithm 2.

The call of `separate` in line 6 of Algorithms 2 and 5 is now completely the same in both algorithms. Since each sub-component returned by `separate` must be decomposable, we check in line 7 of Algorithm 5 if one of these sub-components is already known to be undecomposable. If so, we abort and choose an alternative separator. Otherwise, we try to decompose the sub-components, which is done in the procedure `decompSub` in line 8. The resulting hypertree decompositions are returned in `Subtrees` if the decomposition was successful for all sub-components; otherwise, `Subtrees` $= \emptyset$. Note that there must be at least one sub-component, i.e., `Subtrees` cannot be empty in the case of success. Otherwise, a trivial decomposition of the current component would have been computed in lines 1 to 4 of Algorithm 4. The construction of a new hypertree node is then performed in lines 10 and 11 completely analogous to lines 16 and 17 of Algorithm 2.

Finally, let us consider the procedure `decompSub` defined in Algorithm 6, which contains the functionality of lines 8 to 15 of Algorithm 2. In line 2 we loop through all components that have to be decomposed and in line 3 we compute the set of vertices which will be the second parameter `Conn` of the recursive call of `decompCov`. In order to guarantee the polynomial runtime, we have then to check in line 4 if the current sub-component is already known to be decomposable. If so, we prune the hypertree in line 5; otherwise, we apply the recursive call in line 7 analogous to line 9 in Algorithm 2. Depending on the result, i.e., whether the sub-component could be decomposed or not, we store it either in `FailSeps` or in `SuccSeps`. Moreover, the constructed hypertree decompositions are stored in `Subtrees` in line 15 analogous to line 13 in Algorithm 2.

It is thus easy to see that our algorithm `det-k-decomp` is a deterministic variant of `k-decomp` defined in Algorithms 1 and 2. This insight of the correctness of `det-k-decomp` is now formulated in the following:

**Lemma 2.** *For any given hypergraph $H$ such that $hw(H) \leq k$, `det-k-decomp` accepts $H$. Moreover, for any $c \leq k$, each $c$-width hypertree-decomposition of $H$ in strong normal form is equal to some witness tree for $H$.*

*Proof.* Let $HD = (T, \chi, \lambda)$ be a $c$-width hypertree decomposition of a hypergraph $H$ in strong normal form, where $c \leq k$. We show that there exists an accepting computation tree for `det-k-decomp` on input $H$ which coincides with $HD$. To this aim, note that by Lemma 9 in [3] there exists an accepting computation tree for `k-decomp` on input $H$ which coincides with $HD$. Thus, it suffices to show that every accepting computation tree for `k-decomp` with corresponding hypertree decomposition in strong normal form is also an accepting computation tree for `det-k-decomp`. This, however, is trivial since `det-k-decomp` tries all computation trees of `k-decomp` whose corresponding hypertree decomposition is in strong normal form. □

**Lemma 3.** *If* `det-k-decomp` *accepts a hypergraph $H$, then $hw(H) \leq k$. Moreover, each witness tree for $H$ is a $c$-width hypertree-decomposition of $H$ in normal form, where $c \leq k$.*

*Proof.* We show that every accepting computation tree for `det-k-decomp` on hypergraph $H$ coincides with some $c$-width hypertree decomposition of $H$ in normal form, where $c \leq k$. To this aim, note that by Lemma 13 in [3] every accepting computation tree for `k-decomp` on input $H$ coincides with some $c$-width hypertree decomposition of $H$ in normal form, where $c \leq k$. Thus, it suffices to show that every accepting computation tree for `det-k-decomp` is also an accepting computation tree for `k-decomp`. This, however, is trivial since `det-k-decomp` tries a subset of all computation trees of `k-decomp`. □

By combining Lemma 2 and Lemma 3 we get:

**Theorem 1.** `det-k-decomp` *accepts a hypergraph $H$ if and only if $hw(H) \leq k$.*

Another important property beside the correctness of `det-k-decomp` is its runtime. We gave already some hints why it runs in polynomial time, but we will now consider this in more detail. For simplicity, we assume in analogy to [7] that the standard set operations can be performed in constant time. Moreover, for better comparability, we ignore the auxiliary procedure `expand` since it is not necessary for the decision problem.

**Theorem 2.** `det-k-decomp` *runs in time $\mathcal{O}(n^{k+1} \min(n, ck)^k m^2)$, where $n$ is the number of hyperedges in the hypergraph and $c$ is the maximum number of incident hyperedges over all hyperedges.*

*Proof.* First note that there are at most

$$\Psi = \sum_{i=1}^{k} \binom{n}{i} = \sum_{i=1}^{k} \frac{n!}{i!(n-i)!}$$

separators each with at most $m$ sub-components. Thus, since we restrict the number of recursive calls by checking `SuccSeps` and `FailSeps` to the number of possible separators and sub-components, the number of recursive calls is bounded by $\mathcal{O}(\Psi m)$. What remains to show is the runtime of a single recursive call. To this aim, note that the number of loops in line 6 of Algorithm 4 is bounded by

$$\Phi = \sum_{i=1}^{k} \binom{\min(n, ck)}{i} = \sum_{i=1}^{k} \frac{\min(n, ck)!}{i!(\min(n, ck) - i)!}$$

Moreover, the number of loops in line 4 of Algorithm 5 is bounded by $n$ and the number of loops in line 2 of Algorithm 6 is bounded by $m$. Thus, the runtime of a single recursive call is bounded by $\mathcal{O}(\Phi n m)$. The overall complexity is therefore $\mathcal{O}(\Psi \Phi n m^2)$, which in turn is bounded by $\mathcal{O}(n^{k+1} \min(n, ck)^k m^2)$. □

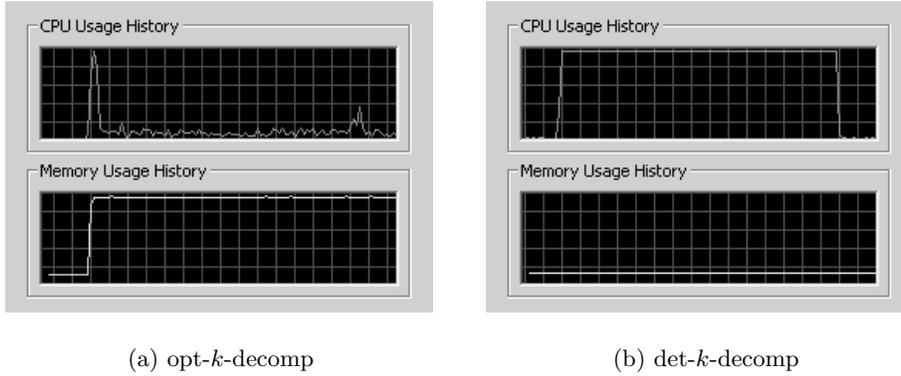

(a) opt-$k$-decomp  (b) det-$k$-decomp

**Fig. 4.** CPU and memory usage when decomposing s298 of the ISCAS89 benchmarks

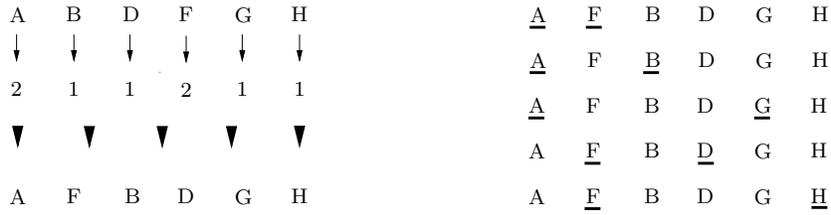

**Fig. 5.** Covering $Conn = \{a, b, e\}$ by $BoundEdges = \{A, B, D, F, G, H\}$

Since we can assume that $ck$ is in many cases much smaller than $n$, we have shown that det-$k$-decomp has a better runtime complexity than opt-$k$-decomp, which runs in time $\mathcal{O}(n^{2k} m^2)$ [7]. Moreover, note that similar to opt-$k$-decomp, the runtime may be significantly smaller in practice. Another important feature of det-$k$-decomp is its low memory usage compared to opt-$k$-decomp. For example, the CPU and memory usage of both algorithms when applied to example s298 of the ISCAS89 benchmark suite can be seen in the screenshot of the Microsoft Windows Task Manager in Fig. 4. Example s298 describes a simple circuit hypergraph consisting of 133 hyperedges and 139 vertices. In Fig. 4 (a), we see that the memory usage suddenly increases to its maximum when applying opt-$k$-decomp; therefore, the system resources are busy with memory swapping such that there are only minimal resources available for solving our decomposition problem. After an hour opt-$k$-decomp was still trying to solve the problem without success. In contrast in Fig. 4 (b), there is an unrecognizably small usage of memory when applying det-$k$-decomp; therefore, the full CPU power can be used for our decomposition problem, which was successfully solved within 90 seconds. We present more experimental results in Section 5, which demonstrate that det-$k$-decomp significantly outperforms opt-$k$-decomp.

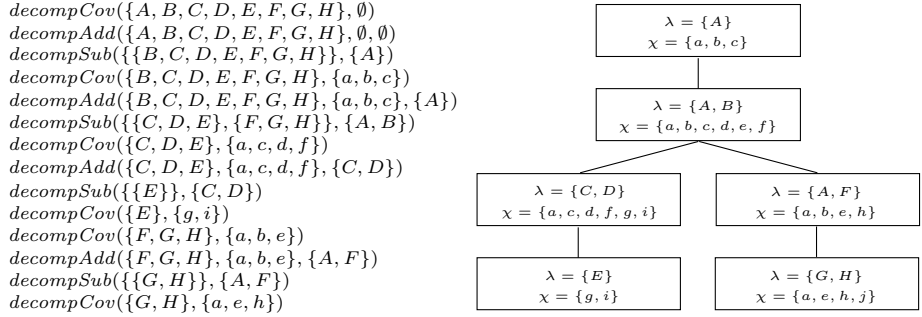

$decompCov(\{A,B,C,D,E,F,G,H\},\emptyset)$
$decompAdd(\{A,B,C,D,E,F,G,H\},\emptyset,\emptyset)$
$decompSub(\{\{B,C,D,E,F,G,H\}\},\{A\})$
$decompCov(\{B,C,D,E,F,G,H\},\{a,b,c\})$
$decompAdd(\{B,C,D,E,F,G,H\},\{a,b,c\},\{A\})$
$decompSub(\{\{C,D,E\},\{F,G,H\}\},\{A,B\})$
$decompCov(\{C,D,E\},\{a,c,d,f\})$
$decompAdd(\{C,D,E\},\{a,c,d,f\},\{C,D\})$
$decompSub(\{\{E\}\},\{C,D\})$
$decompCov(\{E\},\{g,i\})$
$decompCov(\{F,G,H\},\{a,b,e\})$
$decompAdd(\{F,G,H\},\{a,b,e\},\{A,F\})$
$decompSub(\{\{G,H\}\},\{A,F\})$
$decompCov(\{G,H\},\{a,e,h\})$

**Fig. 6.** A procedure call sequence and the corresponding hypertree decomposition

Now, let us come back to our heuristics in the auxiliary procedure `cover`. As we mentioned before, the order of choosing `CovSep` in line 6 of Algorithm 4 has a crucial influence on the performance of `det-`$k$`-decomp`. We obtained our results with a very simple heuristic approach exemplified in Fig. 5 by a decomposition step when decomposing the hypergraph in Fig. 1. Let us assume that `Conn` is given by $\{a,b,e\}$ and `BoundEdges` is given by $\{A,B,D,F,G,H\}$. The task of procedure `cover` is now to compute subsets of `BoundEdges` that cover `Conn` such that at least all minimal covers are considered. Recall that the minimal covers are sufficient because of Lemma 1. We do this in the following way: at first we assign to each hyperedge in `BoundEdges` a weight which is the number of vertices in `Conn` it contains. This can be seen on the left of Fig. 5 where the weight of $A$ is 2 since $A \cap$ `Conn` $= \{a,b\}$, the weight of $B$ is 1 since $B \cap$ `Conn` $= \{e\}$, etc. Then we order the hyperedges according to their weight as it can be seen in the last line on the left of Fig. 5. Finally, we choose the separators based on this ordering in the following way: we consider the hyperedges from left to right and select a hyperedge which covers a vertex in `Conn` that is not covered by other hyperedges selected so far. On the right in Fig. 5, the selection of the five separators $\{A,F\}$, $\{A,B\}$, $\{A,G\}$, $\{F,D\}$, and $\{F,H\}$ is shown. For example, the first one is selected by choosing $A$ which covers $a$ and $b$ and $F$ which covers $e$; the second one is selected by choosing $B$ instead of $F$ to cover $e$; etc. All five separators are computed in this way by choosing covering hyperedges from the left to the right until all vertices in `Conn` are covered. The first separator returned by `cover` is $\{A,F\}$, the second one is $\{A,B\}$, etc. A complete procedure call sequence based on the above heuristics and the corresponding decomposition of the hypergraph in Fig. 1 is now shown in Fig. 6.

In addition to this heuristic approach, we use also some kind of randomization in our algorithm. In particular, the hyperedges read from the input file are randomly ordered. This guarantees that the results are independent from a particular representation of the hypergraph in the input file. However, this also means that `det-`$k$`-decomp` has to be applied several times for evaluation purposes since the results and execution times may differ in each run. Of course, this randomization step can also be removed to take the input order of hyperedges and vertices into account which sometimes improves the obtained results.

## 5 Experimental Results

In this section, we present our experimental results of the algorithm det-$k$-decomp when applied to three different classes of hypergraphs. We compare these results with the results obtained from evaluations of opt-$k$-decomp (which can be downloaded from http://www.deis.unical.it/scarcello/Hypertrees/) and McMahan's heuristic approach. The latter one is based on *Bucket Elimination (BE)* with ordering heuristics *maximum-cardinality search*, *min-fill*, and *min-induced width* [8]. Our evaluations were performed on two different machines since opt-$k$-decomp is only available as Microsoft Windows executable. So opt-$k$-decomp was evaluated under Microsoft Windows 2000 on a 2.4GHz Intel Pentium 4 processor with 512MB main memory and both det-$k$-decomp and BE were evaluated under SuSe Linux 9.2 on a 2.2GHz Intel Xeon processor (dual) with 2GB main memory. Note that the different memory sizes have no relevant influence on the results since the memory usage of det-$k$-decomp and BE is far less than 512MB when applied to our examples. We chose the Intel Xeon as reference machine for normalizing the execution times.

Since BE and det-$k$-decomp use some kind of randomization, we applied them five times to each example in order to obtain representative results. From these five runs we selected the minimal width and computed the average runtime. Moreover, we defined a timeout of 3600 seconds for each run. Since we consider the width as the most important evaluation criterion and not the runtime, we set $k$ always to the smallest value for which we obtained a solution from det-$k$-decomp within our timeout.

The first class of our benchmarks are from DaimlerChrysler and consist of hypergraphs extracted from adder and bridge circuits, the NewSystem examples, and a model of a jet propulsion system [6, 1]. The results when applying our three algorithms to theses hypergraphs are shown in Table 1. The first row in each line contains the name of the example and its size, i.e., the number of hyperedges (atoms) and the number of vertices (variables). The second row contains the hypertree-width if known, i.e., the optimal result. Then there are two rows for each algorithm; the first one contains the minimal width and the second one the average runtime in seconds. It is easy to see that det-$k$-decomp outperforms both $k$-decomp and BE on the DaimlerChrysler examples.

The second class of benchmarks are hypergraphs extracted from two dimensional grids [1]. The advantage of these examples is that their hypertree-width is known by construction. Although we can easily construct an optimal decomposition of these examples by hand, it is seemingly very hard for our algorithms as can be seen in Table 2. Note that opt-$k$-decomp cannot solve any of them within our timeout of one hour. Moreover, note that the runtime of det-$k$-decomp is very high for larger examples. One reason for this may be that we set $k$ to the smallest value for which we obtain a decomposition within an hour. If we increase $k$, then the runtime will decrease in many cases.

| **Instance** *(Atoms / Variables)* | **Min** | **opt-$k$-decomp** | | **BE** | | **det-$k$-decomp** | |
|---|---|---|---|---|---|---|---|
| | | *Width* | *Time* | *Width* | *Time* | *Width* | *Time* |
| **adder_15** (76 / 106) | **2** | **2** | 2 | **2** | 0 | **2** | 0 |
| **adder_25** (126 / 176) | **2** | **2** | 20 | **2** | 0 | **2** | 0 |
| **adder_50** (251 / 351) | **2** | — | — | **2** | 0 | **2** | 0 |
| **adder_75** (376 / 526) | **2** | — | — | **2** | 0 | **2** | 0 |
| **adder_99** (496 / 694) | **2** | — | — | **2** | 1 | **2** | 0 |
| **bridge_15** (137 / 137) | **2** | **2** | 9 | **3** | 0 | **2** | 0 |
| **bridge_25** (227 / 227) | **2** | **2** | 69 | **3** | 0 | **2** | 0 |
| **bridge_50** (452 / 452) | **2** | **2** | 1105 | **3** | 1 | **2** | 0 |
| **bridge_75** (677 / 677) | **2** | — | — | **3** | 1 | **2** | 0 |
| **bridge_99** (893 / 893) | **2** | — | — | **3** | 2 | **2** | 1 |
| **NewSystem1** (84 / 142) | **3** | — | — | **3** | 0 | **3** | 0 |
| **NewSystem2** (200 / 345) | **3** | — | — | **4** | 0 | **3** | 0 |
| **NewSystem3** (278 / 474) | — | — | — | **5** | 1 | **4** | 0 |
| **NewSystem4** (418 / 718) | — | — | — | **5** | 2 | **4** | 0 |
| **atv_partial_system** (88 / 125) | **3** | — | — | **3** | 0 | **3** | 0 |

**Table 1.** DaimlerChrysler benchmarks

| **Instance** *(Atoms / Variables)* | **Min** | **opt-$k$-decomp** | | **BE** | | **det-$k$-decomp** | |
|---|---|---|---|---|---|---|---|
| | | *Width* | *Time* | *Width* | *Time* | *Width* | *Time* |
| **grid2d_10** (50 / 50) | **4** | — | — | **5** | 0 | **4** | 0 |
| **grid2d_15** (112 / 113) | **6** | — | — | **8** | 0 | **6** | 3 |
| **grid2d_20** (200 / 200) | **7** | — | — | **12** | 0 | **7** | 3140 |
| **grid2d_25** (312 / 313) | **9** | — | — | **15** | 3 | **10** | 2000 |
| **grid2d_30** (450 / 450) | **11** | — | — | **19** | 7 | **13** | 1566 |
| **grid2d_35** (612 / 613) | **12** | — | — | **23** | 15 | **15** | 1905 |
| **grid2d_40** (800 / 800) | **14** | — | — | **26** | 28 | **17** | 2530 |
| **grid2d_45** (1012 / 1013) | **16** | — | — | **31** | 51 | **21** | 2606 |
| **grid2d_50** (1250 / 1250) | **17** | — | — | **33** | 86 | **24** | 2786 |
| **grid2d_60** (1800 / 1800) | **21** | — | — | **41** | 204 | **31** | 2984 |
| **grid2d_70** (2450 / 2450) | **24** | — | — | **48** | 474 | **42** | 2161 |
| **grid2d_75** (2812 / 2813) | **26** | — | — | **48** | 631 | **45** | 2881 |

**Table 2.** Grid2D benchmarks

| **Instance** *(Atoms / Variables)* | **Min** | **opt-$k$-decomp** | | **BE** | | **det-$k$-decomp** | |
|---|---|---|---|---|---|---|---|
| | | *Width* | *Time* | *Width* | *Time* | *Width* | *Time* |
| **s27** (13 / 17) | **2** | **2** | 0 | **2** | 0 | **2** | 0 |
| **s208** (104 / 115) | ≥ **3** | — | — | **7** | 0 | **6** | 0 |
| **s298** (133 / 139) | ≥ **3** | — | — | **5** | 0 | **4** | 462 |
| **s344** (175 / 184) | ≥ **3** | — | — | **7** | 0 | **5** | 730 |
| **s349** (176 / 185) | ≥ **3** | — | — | **7** | 0 | **5** | 4 |
| **s382** (179 / 182) | ≥ **3** | — | — | **5** | 0 | **5** | 722 |
| **s386** (165 / 172) | — | — | — | **8** | 1 | **7** | 1824 |
| **s400** (183 / 186) | ≥ **3** | — | — | **6** | 0 | **5** | 273 |
| **s420** (212 / 231) | ≥ **3** | — | — | **9** | 0 | **8** | 454 |
| **s444** (202 / 205) | ≥ **3** | — | — | **6** | 0 | **5** | 385 |
| **s510** (217 / 236) | ≥ **3** | — | — | **23** | 1 | **20** | 2082 |
| **s526** (214 / 217) | ≥ **3** | — | — | **8** | 1 | **7** | 1715 |
| **s641** (398 / 433) | — | — | — | **7** | 1 | **7** | 1611 |
| **s713** (412 / 447) | — | — | — | **7** | 1 | **7** | 1800 |
| **s820** (294 / 312) | ≥ **3** | — | — | **13** | 3 | **12** | 2846 |
| **s832** (292 / 310) | ≥ **3** | — | — | **12** | 3 | **11** | 2575 |
| **s838** (422 / 457) | ≥ **3** | — | — | **16** | 1 | **15** | 2046 |
| **s953** (424 / 440) | ≥ **3** | — | — | **40** | 8 | — | — |
| **s1196** (547 / 561) | — | — | — | **35** | 11 | — | — |
| **s1238** (526 / 540) | — | — | — | **34** | 13 | — | — |
| **s1423** (731 / 748) | — | — | — | **18** | 3 | — | — |
| **s1488** (659 / 667) | — | — | — | **23** | 18 | — | — |
| **s1494** (653 / 661) | — | — | — | **24** | 19 | — | — |
| **s5378** (2958 / 2993) | — | — | — | **85** | 141 | — | — |

**Table 3.** ISCAS89 benchmarks

Finally, the third class of benchmarks are hypergraphs extracted from circuits of the ISCAS89 (International Symposium on Circuits and Systems) benchmark suite. The ISCAS benchmarks are again examples from practice. However, they are obviously much more difficult to solve than the DaimlerChrysler examples as can be seen in Table 3. Note that the widths obtained from `det-`$k$`-decomp` are not much smaller than those from `BE` although it needs much more time.

In summary, we can conclude that `det-`$k$`-decomp` significantly outperforms `opt-`$k$`-decomp` on all classes of benchmarks we have tested. `opt-`$k$`-decomp` terminates within our timeout only for very small and simple examples. In contrast, `det-`$k$`-decomp` terminates also for larger examples with results comparable to or even better than those computed by `BE`. Thus, if the width is the most important criterion and not the runtime, `det-`$k$`-decomp` also outperforms `BE` as long as the hypergraphs are not too large and too complicated.

## 6 Conclusion

We have presented the new algorithm `det-`$k$`-decomp` for constructing hypertree decompositions of hypergraphs. This algorithm results from the alternating algorithm $k$`-decomp` introduced by Gottlob et al. [3] when replacing the *guess and check*-part by a search procedure. We have evaluated `det-`$k$`-decomp` by a series of benchmark examples from industry and academics and compared it to the currently best hypertree decomposition algorithms. Our results have shown that `det-`$k$`-decomp` performs very well; in particular, it significantly outperforms the algorithm `opt-`$k$`-decomp`. Future work is to further restrict the search space by theoretical results and to improve the heuristics used to accelerate the search process.

## References

1. Georg Gottlob, Tobias Ganzow, Nysret Musliu, and Marko Samer. A CSP hypergraph library. Technical Report DBAI-TR-2005-50, Institute of Information Systems (DBAI), TU Vienna, 2005.
2. Georg Gottlob, Nicola Leone, and Francesco Scarcello. On tractable queries and constraints. In *Proceedings of 10th International Conference on Database and Expert System Applications (DEXA)*, volume 1677 of *Lecture Notes in Computer Science*, pages 1–15. Springer-Verlag, 1999.
3. Georg Gottlob, Nicola Leone, and Francesco Scarcello. Hypertree decompositions and tractable queries. *Journal of Computer and System Sciences*, 64(3):579–627, 2002.
4. Georg Gottlob, Nicola Leone, and Francesco Scarcello. Robbers, marshals, and guards: Game theoretic and logical characterizations of hypertree width. *Journal of Computer and System Sciences*, 66(4):775–808, 2003.
5. Peter Harvey and Aditya Ghose. Reducing redundancy in the hypertree decomposition scheme. In *Proceedings of 15th International Conference on Tools with Artificial Intelligence (ICTAI)*, pages 474–481. IEEE Computer Society, 2003.
6. Thomas Korimort. *Constraint Satisfaction Problems – Heuristic Decomposition*. PhD thesis, Institute of Information Systems (DBAI), TU Vienna, April 2003.
7. Nicola Leone, Alfredo Mazzitelli, and Francesco Scarcello. Cost-based query decompositions. Proceedings of the 10th Italian Symposium on Advanced Database Systems (SEBD), 2002.
8. Benjamin McMahan. Bucket eliminiation and hypertree decompositions. Implementation report, Institute of Information Systems (DBAI), TU Vienna, 2004.
9. Francesco Scarcello, Gianluigi Greco, and Nicola Leone. Weighted hypertree decompositions and optimal query plans. In *Proceedings of the 23rd Symposium on Principles of Database Systems (PODS)*, pages 210–221. ACM, 2004. To appear in JCSS.

# A APPENDIX

**Proposition 1.** *The algorithm $k$-decomp presented in Algorithms 1 and 2 is a notational variant of the algorithm $k$-decomp presented in [3].*

*Proof.* First note that $k$-decomp returns NULL if it rejects and it returns a $c$-width hypertree decomposition if it accepts, where $c \leq k$. Algorithm 1 represents the main procedure which calls the recursive procedure $k$-decomposable described in Algorithm 2. The names of these procedures are the same as in their original presentation in [3]. The first parameter of $k$-decomposable, however, is a set of edges instead of a set of vertices. In particular, we use the set $edges(C_R)$ of hyperedges containing some vertices in the component $C_R$ instead of the component $C_R$ itself. Both variants are equivalent since $C_R = \bigcup edges(C_R) \setminus \bigcup R$ can be computed from $edges(C_R)$ and $R$. So the first parameter of $k$-decomposable is the set of hyperedges representing the current component and the second parameter is the set of $\lambda$-labels of the parent node. Note that we use the term *separator* for a set of hyperedges chosen as $\lambda$-labels since they separate the subcomponents.

Now, let us compare Algorithm 2 with its original formulation in [3]. The *guess*-step of Separator in line 1 is obviously the same. For the *check*-step in lines 2 to 4, it is easy to see that both conditions are equivalent to their original formulation. Line 6 describes the computation of the set $\mathcal{C}$ of components in [3]. This computation is now hidden in the procedure separate and its result Components is a set of sets of hyperedges such that for each $C \in \mathcal{C}$ there exists Comp $\in$ Components such that $\bigcup \text{Comp} \setminus \bigcup \text{Separator} = C$. The remainder of Algorithm 2 contains the recursive call as in [3] and the construction of the hypertree decomposition. In particular, the recursive call itself is given in line 9, where we loop through all components. If the decomposition of a component is rejected, we reject in line 11; otherwise, we store the hypertree decomposition of the component in Subtrees in line 13. Finally, a new hypertree node is constructed in line 17 with Separator as $\lambda$-labels and Subtrees as subtrees; the $\chi$-labels are computed according to line 16. Note that the $\chi$-labels computed in this way do not necessarily satisfy the third normal-form condition. However, this condition can be easily satisfied by a simple post-processing step.

So we have demonstrated the equivalence between our notational variant of the algorithm $k$-decomp and its original formulation in [3]. □